\documentclass[useAMS,usenatbib]{mn2e}
\usepackage{times}
\usepackage{epsfig}
\usepackage{amssymb}

\def\be{\begin{equation}}
\def\ee{\end{equation}}
\def\bc{\begin{center}}
\def\ec{\end{center}}
\def\beq{\begin{eqnarray}}
\def\eeq{\end{eqnarray}}

\def\ergs{\rm erg\ s^{-1}}

\title[Ultimate synchrotron cutoff]{Ultimate synchrotron cutoff in gamma-ray spectra of blazars 
as a signature of the converter mechanism}

\author[Boris E. Stern and Yana Y. Tikhomirova]
{Boris~E.~Stern$^{1,2}$\thanks{E-mail:stern@bes.asc.rssi.ru, jana@bes.asc.rssi.ru}
and Yana~Y.~Tikhomirova$^{2}$\footnotemark[1] \\
$^{1}$Institute for Nuclear Research, Russian Academy of Sciences,
Prospekt 60-letiya Oktyabrya 7a, Moscow 117312, Russia\\
$^{2}$Astro Space Centre, Lebedev Physical Institute,
Profsoyuznaya 84/32,  Moscow 117997, Russia}
\begin{document}
\date{Accepted, Received}
\pagerange{\pageref{firstpage}--\pageref{lastpage}} \pubyear{2007}
\maketitle

\begin{abstract}
There is a robust upper limit on the energy of synchrotron 
radiation in high-energy astrophysics: $ \sim m_{\rm e} c^2 /\alpha$, where $\alpha = 1/137$ is the fine 
structure constant and the value refers to the 
comoving frame of the fluid. This is the maximal energy of 
synchrotron photons which can be emitted by an electron having an arbitrarily high initial energy after it turns 
by angle $\sim \pi$ in the magnetic field. This upper limit can be naturally reached if the converter mechanism contributes to the jet radiation and can be imprinted in spectra of some blazars as a cutoff or a dip in the GeV range. We use numerical 
simulations to probe the range of parameters of a radiating jet where the ultimate synchrotron cutoff appears. We reproduce the variety of 
spectra depending on the source luminosity and on the scale of the emission site. We also compare our results with the EGRET blazar spectra in order to illustrate that agreement is possible but still not statistically significant.
The predicted feature, if it exists, should be observed by {\it Fermi} in spectra of some blazars.

\end{abstract}

\section{introduction}

 The issue of maximal possible synchrotron radiation energy in astrophysics (Guilbert, Fabian \& Rees 1983) was discussed 
in various frameworks: the extreme Fermi acceleration of electrons at shocks (e.g. Ghisellini 1999) and the converter mechanism (Derishev 2003; Stern 2003). In the extreme acceleration scenario the electron energy is limited by the condition that the Larmor gyration time, being the shortest available acceleration time, is shorter than the synchrotron cooling time.
In the converter mechanism, a new high-energy electron appears in the relativistic jet as a result of photon-photon pair 
(or photomeson) production and its initial energy can be arbitrarily high in the fluid frame. However, its initial direction in the jet frame is opposite to the jet bulk motion. 
Therefore, its synchrotron radiation at the first 
moment is Doppler deboosted in the external reference frame. When the electron has turned around, the emitted synchrotron photons become Doppler boosted. When it gyrates in the magnetic field by the angle $\pi$, it cools down below the limiting Lorentz 
factor (in the comoving frame) $\gamma_{\rm m} \sim 10^8 B^{-1/2}$, where $B$[G] is the magnetic field strength.
In the extreme acceleration scenario it is approximately the same and in  both cases the maximal 
comoving energy of synchrotron photons does not depend on the magnetic field: $\varepsilon_{\rm m} \sim 1/\alpha$ (i.e. $ \sim 10^2 $MeV), hereafter $\varepsilon = E/m_{\rm e} c^2$ is the photon energy in electron rest mass units. In the case of blazars, this cutoff is blueshifted by factor up to $2\Gamma$, where $\Gamma$ is the bulk Lorentz factor of the jet. Observations of the superluminal proper motion in blazars imply that a typical Lorentz factor is about 
$\Gamma \sim 10-20$ (see Jorstad et al. 2001). In some cases $\Gamma \approx 40$ 
and its value in the emission region could be even higher, when taking into account possible deceleration of the jet (see Stern \& Poutanen 2008, hereafter SP08). Therefore, the ultimate synchrotron cutoff in the blazar spectra should be expected in the GeV range, maybe slightly above 10 GeV in some extreme cases. This feature should be seen as a spectral dip with a continuation of the spectrum to higher energies 
as a result of Comptonization of the external or synchrotron radiation.

In which physical context can the observable synchrotron cutoff in GeV range appear?
The main mechanism of particle acceleration, which is responsible for the gamma-ray 
emission of blazars is still unknown. There are at least three possibilities 
which are discussed:

 (i) {\it Diffusive acceleration of electrons} by various plasma perturbations 
or turbulence (Fermi II mechanism). The ultimate acceleration in this case would imply 
that the relative bulk velocities of the fluid at the gyroradius scale  are relativistic (in the conditions of examples 
considered here the maximal gyroradius  $r_{\rm m} \sim 10^{-5} - 10^{-4} R_{\rm j}$, where $R_{\rm j}$ is the jet radius). This is scarcely possible.

 (ii) {\it Shock acceleration of charged particles} (Fermi I mechanism). It can work at 
internal shocks in the jet  or at the shear layer between the jet 
and the external environment. In this case, acceleration rate can 
be much faster, especially at the shear layer, where an electron entering the jet from the external 
environment can gain factor $\Gamma^2$ in energy (provided that the thickness of the transition layer is 
less than the gyroradius). Therefore, the particle spectra can extend up to $\gamma_{\rm m}$. 
The issue is how much power can be emitted by the particles near the cutoff.
Simulations of shock acceleration (Niemiec \& Ostrowski 2006; Ostrowski 2008) demonstrate a large variety of resulting particle spectra depending on the geometry of the magnetic field and the power spectrum of its perturbations. For some parameters the spectra are very hard. On the other hand, in these simulations the radiative energy losses are neglected and it is difficult to conclude whether they extend close to   $\gamma_{\rm m}$. 
In the shear layer scenario, an electron having the energy above $\gamma_{\rm m}$  cannot penetrate into the jet deeper than 
 $r_{\rm m}$. Therefore, the energy budget for extreme acceleration at the shear layer is restricted by the bulk energy of a thin outer layer of the jet.

 (iii) {\it Converter mechanism}, which can take a 
form of a runaway photon breeding in relativistic jets (Derishev et al. 2003; Stern 2003; Stern \& Poutanen 2006). 
The photon breeding mechanism is probably the most efficient  way of dissipation of the relativistic jet energy into gamma-rays. The mechanism should work if the active galactic nucleus (AGN) is 
sufficiently powerful  (above $L_{\rm d} \sim 10^{42} - 10^{44}\ergs$, depending on the scale of the emission site) and the if jet is sufficiently 
fast  (the Lorentz factor above $\Gamma \sim 10$, see SP08).
When photon breeding mechanism operates, the instant energy gain at photon -- electron-positron conversion in 
the jet  can reach $4 \Gamma^2$ and the ultimate energy $\gamma_{\rm m}$ can be 
easily reached. Stern \& Poutanen (2006), studying the photon 
breeding regime in jets, showed that  
the energy of particles through the breeding cycle grows till the 
Lorentz factor of produced pairs reaches  $\gamma_{\rm m}$ in the jet frame. SP08 performing numerical simulations of the photon breeding 
have obtained some hard ($\beta \sim 2$) spectra at a moderate luminosity of the source, where the ultimate synchrotron cutoff is clearly seen (hereafter we define the photon spectral index $\beta$ as  ${\rm d}N/{\rm d}\varepsilon\propto \varepsilon^{-\beta}$).
  As discussed above, the possibility to reproduce such cutoff with other mechanisms is uncertain at best.
  Therefore, the observation of such a feature would be an 
argument in favor of the converter mechanism as a source of the jet radiation.

 Existing high-energy data supplied mainly by EGRET (30 MeV -- 20 GeV energy range),  do not reveal such a cutoff
at a significant level. Nevertheless, there are some indications of a nontrivial shape of
the blazar spectra in the high-energy range, more complicated than a single high-energy hump. 
Such an indication has been found  by Costamante, Aharonian \& Khangulyan (2007) in the spectrum of BL Lac 
PKS 2155--304. This is one of five objects, which has been detected by both EGRET and Cherenkov arrays: the
combination of these two data sets imply that there is a double humped structure in the ten MeV  -- TeV range with a depression around 10 GeV.   

  This study is motivated by the launch of {\it Fermi} (former {\it GLAST}). This instrument 
covering 200 MeV -- 200 GeV range is perfect for detection of the feature we 
study (see also discussion in Costamante et al. 2007).

In this work we perform a series of simulations 
in order to outline the variety of conditions, 
where the ultimate synchrotron cutoff can be observed, and to demonstrate the variety of spectral 
shapes in the MeV -- TeV range produced by the photon breeding mechanism. The 
problem setup, described in Section 2, is similar to that of SP08. In the same section we present results of numerical 
simulations and discuss them. In Section 3 we compare our results with the EGRET blazar data in order to illustrate that the existing data have slightly insufficient accuracy to make significant conclusions about the structure of blazar spectra in the GeV range.
In section 4 we discuss the perspectives of observations.

\section{The numerical simulations}
  
  We have  performed Monte-Carlo simulations of the jet radiation in the photon 
breeding regime using the Large Particle Monte-Carlo Code (LPMC) for particle 
propagation and interactions  and a two dimensional ballistic model for the 
jet dynamics. The problem setup and the technique of numerical simulations 
are described in  SP08, 
with details given by Stern \& Poutanen (2006). LPMC code is described 
in Stern et al. (1995). 

Briefly, the problem setup can be summarized as follows: 
\begin{enumerate}
 \item  {\it The jet} is represented as a cylinder of radius $R_{\rm j}$ and length $\Delta z = 20 R_{\rm j}$ split into array 64$\times$20 of 
concentric shells and slices, where each cell can independently decelerate by 
exchanging 4-momentum with interacting photons. The internal pressure is neglected. The initial Lorentz factor $\Gamma $ at the inlet is constant. The magnetic field has toroidal geometry. 

\item {\it Initial state.} A simulation starts with a uniform jet of a constant Lorentz factor and a small amount of seed high-energy photons. 

\item {\it The external environment} is kept at rest (see SP08 for the corresponding condition) and has the magnetic field perpendicular to the jet with random orientation in azimuthal direction. 

 \item {\it Emission sites.} SP08 studied three possible emission sites: at the scale of the broad emission line region ($R \sim 10^{17}$cm, model A), the parsec scale (model B) and small scale ($R \sim 30 - 100 R_{\rm g} \sim 10^{15}$ cm, where $R_{\rm g}$ is the gravitational radius, model C). Here we consider models A and C.

 \item {\it The external radiation.} For model A we assume two-component radiation field, which includes direct radiation of the accretion disc (the multicolor blackbody spectrum and the X-ray tail) and the isotropic radiation with the spectrum extending
from the UV (lines and scattered disc radiation) to the far IR (dust radiation 
from a few parsec scale and a possible contribution of synchrotron radiation). 
The energy density of the isotropic component is 0.05 of that
of the direct disc radiation. The spectrum of the isotropic component is 
parametrised as a power law ${\rm d}N/{\rm d}\varepsilon \propto \varepsilon ^{-1.4}$ in the 
range $10^{-8} < \varepsilon <  T_{\rm max}$, where $T_{\rm max} = 0.5 \times 10^{-5}$ is the maximal temperature of the accretion disc in $m_{\rm e} c^2$ units. 
Model  C represents a smaller distance scale: the vicinity of the accretion disc at tens -- hundreds gravitational radii. At such a distance the photon 
breeding does 
not require a separate isotropic component of the soft background radiation, because
the accretion disc supplies some photons at sufficiently large 
angles. In the case of model C, the picture of radiation field is more certain than that in model A: the temperature 
of the disc varies as $T = T_{\rm max} (r/r_{\rm min})^{-3/4}$,  the luminosity is 
distributed as ${\rm d}L/{\rm d}r \propto r^{-2}$, the angular distribution of the emitted 
photons flux is described by the Lambert law ${\rm d}F/{\rm d\ cos} \theta \propto {\rm cos}\ \theta$.
The disc radiation field was precomputed as a function of the distance from the central 
engine in the range $( 30 - 100) R_{\rm g}$.
\end{enumerate}

\begin{figure}
\begin{center}
\leavevmode \epsfxsize=7.0cm \epsfbox{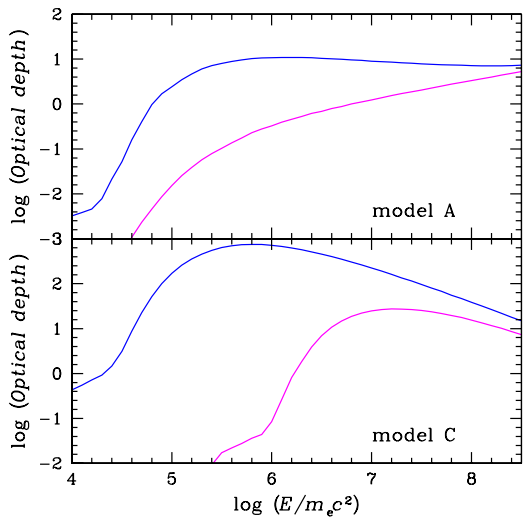}
\end{center}
\caption{ \label{fig:opac}
The photon-photon opacity at the photon path $R_j$ depending on the energy 
of photons moving in the counter-jet (upper curves for both models)  and 
the jet direction (lower curves for both models). The absolute value of the opacity corresponds to the disc luminosity $2 \times 10^{44}\ergs$ for model A and $2.5 \times 10^{43} \ergs$ for model C.} 
\end{figure}

 In both models, the spectrum of the disc radiation includes a nonthermal X-ray tail described by the power law with photon index $\beta = 2$  extending up to 100 keV with the total luminosity 0.07 of that of the thermal component. Such X-ray emission exists in many AGNs. For model C the X-ray component is important, because it provides a sufficient opacity  for photons of energy $\varepsilon \sim 10^4$ which are otherwise too soft to interact with the thermal disc photons (see Fig. \ref{fig:opac}). 

The main difference between the two models can be expressed in terms of 
dependence of the photon-photon opacity on the direction. Fig. \ref{fig:opac} shows
the opacity  for photons moving along the jet and in the opposite direction. 
One can see that in the case of model C, the anisotropy in the opacity is higher 
than that in model A.

We have chosen these two models as idealised extreme versions of external 
radiation: the emission of a point-like disc plus an isotropic background (model A) and the radiation of a geometrically large accretion disc only (model C). Actually the soft radiation field should be their superposition. Particularly, model C should also include some scattered radiation, which we neglect.

The relevance of each model for different classes of objects is not evident.
For the operation of the photon breeding in model C, a lower luminosity 
of the accretion disc is required than in  model A: $L_{\rm d} \sim 10^{42} R_{15} \ergs$ versus  $\sim 10^{43} R_{17}\ergs$ ($R_{n}= R / 10^n$cm). In units of compactness (which is proportional to 
$L_{\rm d}/R$), model A has a lower threshold. Model C is probably the only one which can
work in the case of BL Lacs. On the other hand, model C implies that the  acceleration of the jet is fast: 
the Lorentz factor $\Gamma > 20$ should be reached at a few tens $R_{\rm g}$. Otherwise this model does not provide the conditions for the exponential photon breeding process.
In model A the conversion of the jet energy into the radiation is  more efficient.

Other parameters for model A: $\Gamma = 20$, the average distance from the black hole 
$2 \times 10^{17}$ cm, the jet radius $R_{\rm j} =10^{16}$cm, the jet power $L_{\rm j} = L_{\rm d}$, the magnetic energy flux $L_{\rm B} = 0.2 L_{\rm d}$. Note that the jet power  is not important unless nonlinear effects become substantial. The latter take place at $L_{\rm j} \sim 10^{45} \ergs$. 

 Parameters for model C: the  distance between the beginning of the active region and the black hole $R=10^{15}$ cm, gravitational radius (which defines the accretion disc scale) 
$R_{\rm g} = 3 \times 10^{13} $cm, the jet radius $R_{\rm j} = 10^{14}$cm, $L_{\rm B} = 0.25 L_{\rm d}$, $L_{\rm j} =L_{\rm d}$.

We have performed a series of simulations for each model varying the density of
the external soft photon field. This was parametrised through the 
corresponding luminosity of the accretion disc  $L_{\rm d}$. In both cases we have started 
from the luminosity which is slightly above the threshold for the supercritical 
photon breeding and have increased it by steps small enough to trace the evolution of the spectrum. For model A the simulations (A1--A4) correspond to 
$L_{\rm d}= (0.5, 1, 1.4, 2)\times 10^{44} \ergs$.   For the model C (runs C1--C6): 
$L_{\rm d}= (1, 2, 4, 8, 16, 32) \times10^{42} \ergs $.

\begin{figure}
\begin{center}
\leavevmode \epsfxsize=7.0cm \epsfbox{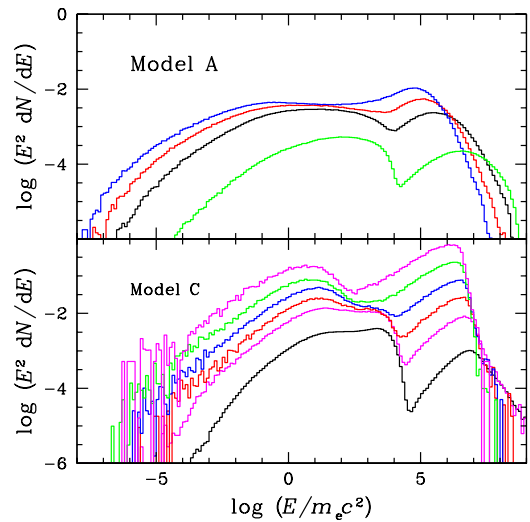}
\end{center}
\caption{ \label{fig:spec}  
Spectra produced by the photon breeding mechanism obtained with 
the numerical simulations. Model A: 
curves from bottom to top correspond to runs A1--A4 
(disc luminosity varies from 
$0.5 \times 10^{44}\ergs$ to $2 \times 10^{44} \ergs$). 
Model C: curves from bottom to top 
correspond to runs C1--C6 
(the disc luminosity varies from $10^{42}\ergs $ to  $3.2\times 10^{43}\ergs$).}
\end{figure}

The resulting efficiency of the jet power conversion into radiation varies from 0.16 (run A1) to 0.6 (run A4) for model A and from 0.05 (run C1) to 0.3 (run C6) for model C. 
The resulting spectra as a function of the disc luminosity for each model are given in Fig \ref{fig:spec}.  

In both cases the ultimate synchrotron component is relatively strong and 
the cutoff is well seen at a low disc luminosity: within factor 
$\sim 3 $ above the threshold of supercritical photon breeding  in the case of 
 model A, and within one order of magnitude in the case of model C.
In real sources, the compactness parameter is probably widely distributed, and  
therefore the number of objects demonstrating the ultimate synchrotron cutoff in
its clear form should be relatively small. At a higher disc luminosity the cutoff
turns into a shallow dip. In both cases, the feature 
is associated with the transition between the synchrotron and the Compton components,
however, the position and the shape of the break are different for the two models.  

 The reasons why the transition between the synchrotron and the Compton components shifts to a lower energy as the compactness increases are as follows:

(i) In the case of model C the high-energy photons ($\varepsilon \sim 1/T_{\rm max}$ produced {\it in the external environment} are removed from the breeding cycle.
Indeed, the photon-photon opacity  across the jet for $\varepsilon =1/T_{\rm max}$ 
can be as large as a few hundred and most of the external high-energy photons are absorbed beyond the jet. Only the photons with $\varepsilon < 5\times 10^4$ can 
penetrate the jet deeply enough (see Fig.\ref{fig:opac}) to support efficiently  the breeding cycle. The synchrotron component induced by  these photons is cut off at 
\begin{equation}
\varepsilon_{\rm s} \sim (\varepsilon \Gamma)^2 {B \over B_{\rm 0}} \Gamma \sim 10 {\big(}{\Gamma \over 30}{\big)}^3 {B \over 10{\rm G}}, 
\end{equation}
where $B_{\rm 0} = 4.4 \times 10^{13}$G. Indeed, one can see a maximum at $\varepsilon \sim 10$ for model C at a high compactness.

(ii) For model A the high-energy photons produced {\it in the jet} are absorbed because of a high opacity produced by the isotropic soft photons. One can see in Fig.\ref{fig:opac} that for model A the opacity of the isotropic background at the path length $\sim 10 R_{\rm j}$ exceeds unity when 
$\varepsilon > 10^6$. The corresponding absorption turnover is seen  
in Fig. \ref{fig:spec}. In this case, the absorption initiates the electromagnetic cascade  inside the jet producing softer pairs and broader spectra without sharp features.

 The variety of spectra we present here is incomplete. For example, when the ratio between the magnetic energy flux and the disc luminosity is increased, the Compton peak becomes lower and the jet radiative efficiency decreases. A larger series of simulations will be presented in a separate publication.
 We also do not discuss the
 observed low-energy synchrotron maximum treated by SP08 as a 
secondary effect: radiation of pairs in heating/cooling balance further 
downstream along the jet.

\section{Egret data}

The EGRET catalog  (Hartman et al. 1999) includes  66 high-confidence
identifications of blazars (BL Lac objects, flat-spectrum radio quasars,
or unidentified flat-spectrum radio sources).

\begin{figure}
\begin{center}
\leavevmode \epsfxsize=7.0cm \epsfbox{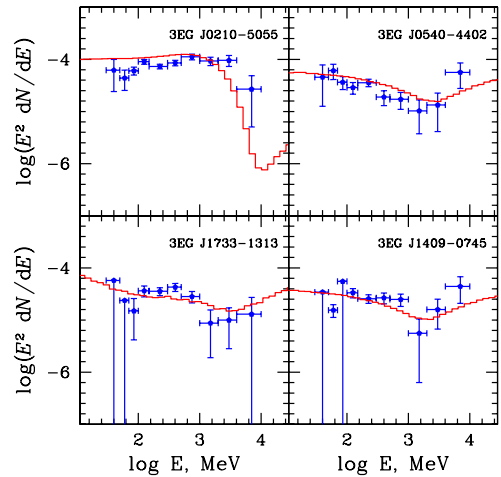}
\end{center}
\caption{ \label{fig:egret}  
Examples of the EGRET spectra sampled from the seven spectra with largest reduced $\chi^2$  of a power-law fit 
resembling some two-component simulated curves from Fig. \ref{fig:spec}.
 None of these examples gives a significant confirmation of the two-componentness because of the large data errors. 
The simulated curves correspond to:  run C1 at $z$=1.003 (3EG J0210--5055), 
the run C4 at $z$=0.902 (3EG 1733--1313),
the run A2 at $z$=0.894 (3EG J0540--4402) and 
the run A2 at $z$=1.494 (3EG J1409--0745).
}
\end{figure}

 The data errors are rather large and 
all spectra of 66 blazars with a few exceptions can be fit with a simple power law at a satisfactory $\chi^2$. In order to illustrate the present accuracy of observations we show four of seven spectra with largest deviations from a power law (reduced $\chi^2 > 2$) in comparison  with our simulated spectra applying their measured redshift (Fig. \ref{fig:egret}). 
In  these cases our spectra give a better description than a power law.  In the case of 
3EG J0210--5055 the cutoff in the simulated curve is at a slightly smaller energy than is indicated by the data, but this could be compensated by a larger Lorentz factor of the jet. 

Of course, none of these examples confirms the existence of the synchrotron turnover in the GeV range: the statistical significance is insufficient.

 An additional evidence of the high-energy synchrotron maximum could be obtained from comparison of the EGRET and the TeV 
spectra for those objects, where both kinds of data exist. At present, there are 
five blazars detected both by EGRET and Cherenkov telescopes. Note that observations in the EGRET and the TeV ranges are  not simultaneous. 

 1. PKS 2155--304. As we already mentioned in the Introduction this case has been studied by Costamante et al. (2007) as a possible example of the synchrotron maximum in the EGRET range.

 2. Mkn 421. It has a comparatively hard spectrum in the EGRET range ($\beta \sim 1.6$) extending up to 10 GeV, which is inconsistent with the high-energy synchrotron component unless the Lorentz factor of the jet is very large $\Gamma > 50$.

 3. 1ES 1959+650. The TeV spectrum is published by Aharonian et al. (2003). The EGRET spectrum is soft  ($\beta \sim 2.5$) with the GeV energy flux  $ \varepsilon^2 {\rm d}N/{\rm d}\varepsilon \sim 10^{-11}$ TeV s$^{-1}$ cm$^{-2}$ while the TeV energy flux corrected for the infrared background absorption is slightly higher. This resembles the case of  PKS 2155--304 and intermediate curves in Fig.2.

 4. 3C 279. In the EGRET range the spectrum has a smooth maximum at 
1 GeV. The spectrum of TeV emission discovered by Teshima et al. (2007) has not been published yet.

5. BL Lacertae. The spectrum is very variable both in the EGRET and the TeV ranges (see Albert et al. 2007), and no clear conclusions can be made because of the lack of simultaneous observations.

The above consideration just illustrates the present observational status.
The data do not reject the hypothesis of the existence of the ultimate 
synchrotron cutoff in some blazar spectra, but cannot confirm it either, because of the insufficient gamma-ray statistics. However, it is clear that even a moderate increase in the statistics will provide sufficient significance in cases like those presented in Fig. \ref{fig:egret}. 

\section{Conclusions and perspectives of observations}

Our simulations demonstrate that the ultimate synchrotron cutoff 
in its clear form is a relatively rare phenomenon. In the framework of the photon breeding mechanism, the cutoff in the GeV range appears when the compactness of the external radiation is just above the threshold of the supercritical photon breeding. When the compactness 
exceeds the threshold by an order of magnitude, the spectrum becomes almost flat or the synchrotron component moves to the softer range (see Fig. \ref{fig:spec}). 
Another tendency, which can be derived from the results, is that the spectral sequence depends on the angular distribution of the external radiation. When the large-angle component of the external radiation in the emission region is relatively strong (e.g. $\sim 0.1$ of the direct disc radiation as in model A) the spectra are smoother and the separation between the 
synchrotron and Compton components is weakly expressed. When the side-on radiation is weak, e.g. it is supplied only by the periphery of the accretion disc, all features are  much sharper.

 The number of blazars with spectra demonstrating the clear ultimate cutoff should be relatively small. However, {\it Fermi} will be able to reveal this kind of spectra if it exists, because it will observe simultaneously a large number of objects. 
The cutoff in the right place is by itself insufficient to conclude that we observe the ultimate synchrotron cutoff:
the latter can also be caused by the photon-photon opacity in the same energy range (see SP08). To be sure that the observed cutoff has a synchrotron nature, one have to observe the continuation of the spectrum above the cutoff in a form of the
higher energy Compton component extending to the TeV range. In principle, {\it Fermi} can
detect the  low-energy part of the Compton maximum at $E \sim 100$ GeV and confirm the two-componentness in this way even without detection of the TeV 
emission by large ground-based detectors.

Although the possibility of the extreme acceleration up to the energy $\gamma_{\rm m}$ in internal shocks is not evident, it is not yet ruled out. Do independent signatures exist of the photon breding mechanism?
A possible one is the correlation between the disc brightness and the spectral shape like those in Fig.  \ref{fig:spec}.

If the total power of the jet is known, then the high radiative efficiency, i.e. the ratio of the gamma-ray luminosity to the total jet power,  
of e.g. more than 10 per cent can give an additional support to the photon breeding hypothesis and argue against the internal shock scenario. In the internal shock mechanism the luminosity budget is limited by 
the {\it internal energy} of the jet, while in the case of the photon breeding a large fraction of the {\it total bulk energy} of the jet  can
be converted into radiation. 

Another possible clue is the compactness threshold:  at a low compactness of the disc radiation, the photon breeding does not work (the threshold for  model C is  $L_{\rm d}/R_{15} \sim (1 - 10)\times 10^{42} \ergs$). Usually it is 
difficult to estimate the distance $R$ from the disc to the emission region, however, sometimes approximate constraints can be obtained using the time variability
or the mass of the black hole ($R$ can not be smaller than a few $R_{\rm g}$).
In this way the photon breeding mechanism can be revealed statistically: sources with the low estimated compactness should be  less efficient and have different spectra. When the photon breeding turns on, the luminosity should increase sharply. Therefore one can expect a kind of bimodality in the efficiency and probably in the high-energy spectra of blazars.

 \section*{Acknowledgments}
 
The work is supported by the Russian Foundation for Basic Research 
(N07-02-00629a) and the Presidium of Russian Academy
of Sciences (Scientific school 2469.2008.2 and ''Origin and
evolution of stars and galaxies'').
We thank Pavel Ivanov, Juri Poutanen and Stephanie Patoir  for useful comments.


\begin{thebibliography}{99}
\bibitem[\protect\citeauthoryear{Aharonian et al.}{2003}]{a03}
Aharonian F. A. et al., 2003, A\&A, 406, L9


\bibitem[\protect\citeauthoryear{Albert et al.}{2007}]{a07}
Albert J. et al., 2007, ApJ, 666, L17



\bibitem[\protect\citeauthoryear{Costamante et al.}{2007}]{co07}
Costamante L., Aharonian F., Khangulyan D., 2007, 
AIP Conf. Proc., 921, 157

\bibitem[\protect\citeauthoryear{Derishev et al.}{2003}]{der03}
Derishev E. V., Aharonian F. A., Kocharovsky V. V.,
Kocharovsky Vl. V., 2003, Phys Rev D, 68, 043003

\bibitem[\protect\citeauthoryear{Ghisellini}{1998}]{ghi98}
Ghisellini G., 1999, Astroph. Lett. \& Comm., 39, 17 

\bibitem[\protect\citeauthoryear{Guilbert et al.}{1983}]{guil}
Guilbert P. W., Fabian A. C.\& Rees, M. J., 1983, MNRAS, 205, 593

\bibitem[\protect\citeauthoryear{Hartmann et al.}{1999}]{egret}
Hartman  R. C. et al., 1999,  ApJS, 123, 79

\bibitem[\protect\citeauthoryear{Jorstad et al.}{2001}]{jor01}
Jorstad S. G., Marscher A. P., Mattox J. R., Wehrle A. E., Bloom S. D., Yurchenko A. V.,
2001, ApJS, 134, 181

\bibitem[\protect\citeauthoryear{Niemiec \& Ostrowski}{2006}]{no06}
Niemiec J., Ostrowski M., 2006, ApJ, 641, 984 

 %
\bibitem[\protect\citeauthoryear{Ostrowski}{2008}]{ost08}
Ostrowski M., 2008, preprint (arXiv:0801.1339)

\bibitem[\protect\citeauthoryear{Stern}{2003}]{st03}
Stern B. E., 2003, MNRAS,  345, 590 (ST03)

\bibitem[\protect\citeauthoryear{Stern et al.}{1995}]{sbss95}
Stern B. E., Begelman M. C., Sikora M., Svensson R., 1995, MNRAS, 272, 291

\bibitem[\protect\citeauthoryear{Stern \& Poutanen}{2006}]{sp06}
Stern B. E., Poutanen J., 2006, MNRAS, 372, 1217 

\bibitem[\protect\citeauthoryear{Stern \& Poutanen}{2006}]{sp08}
Stern B. E., Poutanen J., 2008, MNRAS, 383, 1695 (SP08)

\bibitem[\protect\citeauthoryear{teshima et al.}{2007}]{t07}
Teshima M. et al., 2007, preprint (arXiv0709.1475)


\end{thebibliography}
\end{document}